\documentclass{ceurart}

\usepackage{xcolor}
\usepackage{listings}
\lstdefinestyle{yaml}{
  basicstyle=\ttfamily\small,
  commentstyle=\color{black!55},
  showstringspaces=false,
  columns=fullflexible,
  keepspaces=true,
  breaklines=true,
  breakatwhitespace=true,
  tabsize=2,
  frame=single,
  rulecolor=\color{black!30},
  comment=[l]{\#},
}
\begin{document}

\copyrightyear{2027}
\copyrightclause{Copyright for this paper by its authors.
  Use permitted under Creative Commons License Attribution 4.0
  International (CC BY 4.0).}

\conference{SWAT4HCLS 2027: 18th International Conference on Semantic Web
  Applications and Tools for Health Care and Life Sciences,
  February 01--04, 2027, Basel, Switzerland}

\title{kgsteward: a tool for building, reproducing and maintaining distributed
  knowledge graphs}

\author[1]{Marco Pagni}[%
  orcid=0000-0001-9292-9463,
  email=Marco.Pagni@sib.swiss,
]

\author[1]{Robin Engler}[%
  orcid=0009-0001-7090-6196,
  email=Robin.Engler@sib.swiss,
]

\author[1]{Frédéric Burdet}[%
  orcid=0000-0002-2923-827X,
  email=Frederic.Burdet@sib.swiss,
]

\author[1]{Sébastien Moretti}[%
  orcid=0000-0003-3947-488X,
  email=Sebastien.Moretti@sib.swiss,
]

\author[2,3]{Loïc Le Gregam}[%
  orcid=0009-0008-1993-4918,
  email=loic.legregam@unige.ch,
]

\author[2,3]{Luis Quiros-Guerrero}[%
  orcid=0000-0002-1630-8697,
  email=luis.quiros@rd.nestle.com,
]

\author[4]{Jahn Nitschke}[%
  orcid=0000-0003-3838-0945,
  email=nitschke.jahn@gmail.com,
]

\author[1,5]{Pierre-Marie Allard}[%
  orcid=0000-0003-3389-2191,
  email=pierre-marie.allard@unifr.ch,
]

\author[2,3,6]{Louis-Félix Nothias}[%
  orcid=0000-0001-6711-6719,
  email=louis-felix.nothias@univ-cotedazur.fr,
]

\author[7]{Jack McGoldrick}[%
  orcid=0009-0000-2106-4159,
  email=j.mcgoldrick9@universityofgalway.ie,
]

\author[7]{Ronan M.T. Fleming}[%
  orcid=0000-0001-5346-9812,
  email=ronan.mt.fleming@universityofgalway.ie,
]

\author[8]{Alan Bridge}[%
  orcid=0000-0003-2148-9135,
  email=Alan.Bridge@sib.swiss,
]

\author[1]{Vincent Emonet}[%
  orcid=0000-0002-1501-1082,
  email=Vincent.Emonet@sib.swiss,
]

\author[1]{Tarcisio Mendes de Farias}[%
  orcid=0000-0002-3175-5372,
  email=Tarcisio.Mendes@sib.swiss,
]

\author[1]{Ana-Claudia Sima}[%
  orcid=0000-0003-3213-4495,
  email=Ana-Claudia.Sima@sib.swiss,
]

\author[1]{Mark Ibberson}[%
  orcid=0000-0003-3152-5670,
  email=Mark.Ibberson@sib.swiss,
]

\author[2,3]{Jean-Luc Wolfender}[%
  orcid=0000-0002-0125-952X,
  email=Jean-Luc.Wolfender@unige.ch,
]

\author[1]{Florence Mehl}[%
  orcid=0000-0002-9619-1707,
  email=Florence.Mehl@sib.swiss,
]
\cormark[1]

\address[1]{Vital-IT, SIB Swiss Institute of Bioinformatics,
  Lausanne, Switzerland}
\address[2]{Institute of Pharmaceutical Sciences of Western Switzerland,
  University of Geneva, Geneva, Switzerland}
\address[3]{School of Pharmaceutical Sciences, University of Geneva,
  Geneva, Switzerland}
\address[4]{Department of Biochemistry, Faculty of Science,
  University of Geneva, Science II, Geneva, Switzerland}
\address[5]{Department of Biology, University of Fribourg,
  Fribourg, Switzerland}
\address[6]{Université Côte d'Azur, Institut de Chimie de Nice,
  Campus Valrose, Nice, France}
\address[7]{School of Medicine, University of Galway, Galway, Ireland}
\address[8]{Swiss-Prot Group, SIB Swiss Institute of Bioinformatics,
  Geneva, Switzerland}

\cortext[1]{Corresponding author.}

\begin{abstract}
  Collaborative research projects in life sciences increasingly need to
  integrate private, embargoed consortium data with public reference databases
  in order to reach statistically meaningful interpretations. The Resource
  Description Framework (RDF) is well suited to this task: it facilitates the
  integration of heterogeneous data sources, and allows researchers to keep data
  and their documentation as metadata in the same place, provided the knowledge
  graph itself remains private during the time course of the project.
  Nevertheless, the development and long-term maintenance of a scientific
  knowledge graph remains a challenging, labour-intensive endeavour owing to the
  state of constant flux of most public resources. To tackle this challenge, we
  present \emph{kgsteward}, a Python command-line tool that builds and maintains
  knowledge graphs inside RDF stores from a single, version-controlled
  configuration file. \emph{kgsteward} supports multiple triplestores, keeps the
  local graph up-to-date with its external sources possibly already in RDF, or
  transformed into it on the fly, and uses SPARQL 1.1 UPDATE commands to amend
  further imported RDF on the fly. It can also validate the resulting graph with
  SPARQL queries that double as usage examples for both human users and AI
  agents. \emph{kgsteward} has already been used in several collaborative
  projects at the SIB Swiss Institute of Bioinformatics, and we demonstrate its
  applicability in two real-world international research projects: one that
  builds a library of plant extracts with chemical analyses and associated
  bio-activities, and a second that reconciles public reference resources for
  human metabolic-network reconstruction.
\end{abstract}


\maketitle


\section{Introduction}
\label{sec:intro}

Collaborative research projects generally require using both private consortium
data --- held under restricted access --- and public databases, which allow
incorporating broader context information and improve statistical robustness of
the results. Choosing an appropriate data-management technology for such
projects is not obvious, because the data are heterogeneous, evolve continuously
throughout the project, and are produced and consumed by collaborators with
vastly different levels of technical expertise.

We adopted RDF rather than another technology for four main reasons. First, and
most decisively, much of the reference data such projects depend on is
\emph{already published as RDF}: SIB alone distributes UniProt, Rhea,
SwissLipids, Bgee, OMA and MetaNetX through public SPARQL endpoints, and
comparable resources are available elsewhere. These can be integrated as they
are, an advantage no relational design can offer. Second, RDF is a flexible data
model that facilitates integration in spite of heterogeneity: new sources and
new properties can be added to an existing graph without migrating a schema,
which matters when the data model keeps evolving over the life of a project.
Third, global IRIs make links between resources explicit and durable, so that
private and public data share a single identifier space. Fourth, RDF keeps
documentation (schemas, ontologies, …) and data in the same place, providing AI
agents~\cite{emonetLLMbasedSPARQLQuery2025} enough context to query a graph they
have not seen before. Together, these properties make it considerably easier to
meet the FAIR principles and the data-management-plan obligations imposed by
funding agencies.

These advantages come with practical demands that any long-lived knowledge-graph
project must meet: staying portable across triplestore implementations rather
than locked to one vendor; integrating public and private sources and keeping
them up-to-date as remote databases evolve (UniProt, for example, is released
every two months); rebuilding reproducibly from scratch; and letting each
project partner quickly deploy and extend a common base locally. Manually
implementing these requirements means accumulating fragile, store-specific
scripts that only their author can run, precisely what a consortium of
non-specialists cannot sustain. What is needed instead is a tool centred around
a single configuration file, which can be kept under version control, and from
which anyone can rebuild the graph with a single command.

Here we present \emph{kgsteward}, a Python-based command-line tool designed to
build and maintain knowledge graphs within RDF stores. Its development started
in 2021 and has since been used in several collaborative research projects at
SIB. This paper describes the software and the lessons learned from its
deployment through two contrasting use cases.

\section{Software description}
\label{sec:software}

\begin{figure}[!ht]
  \centering
  \includegraphics[width=\linewidth]{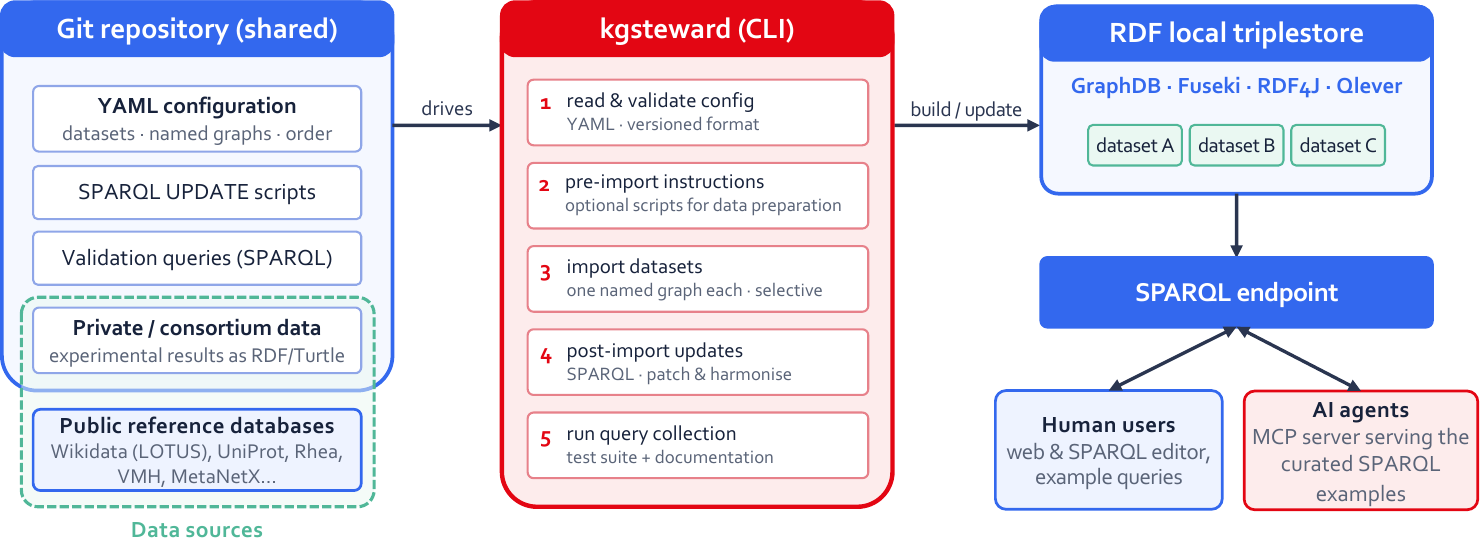}
  \caption{The \emph{kgsteward} execution model. A single configuration,
    ideally version-controlled, drives the build (or update) and validation
    of the local KG.}
  \label{fig:execution-model}
\end{figure}

\emph{kgsteward}~\cite{pagniKGSteward2023} is a CLI tool that manages the
content of an RDF store from a declarative configuration file rather than
through the graphical web interface of a triplestore
(Figure~\ref{fig:execution-model}).

\subsection{Overall principle}
\label{sec:principle}

\emph{kgsteward} maintains the content of an RDF triplestore from a declarative
specification in the form of a plain-text YAML file listing one or more RDF data
sources (hereafter referred to as datasets) from which a local RDF triplestore
(hereafter knowledge graph; KG) is populated. A \textbf{dataset} is the unit of
declaration, data import and change-tracking. Each dataset carries a unique
short name and produces exactly one \textbf{named graph} (also known as a
\enquote{context} in the rdf4j/GraphDB jargon) in the local KG. For each
dataset, the configuration YAML file specifies the data source(s) and the
datasets it depends upon, declared as its \enquote{parents}. Updating a parent
causes its children to be updated in turn. The configuration file, together with
the data sources it refers to, thus constitutes a complete and portable
description of the KG. The config file is typically kept under Git version
control, allowing the construction of the KG to be repeated elsewhere.

\begin{figure}[!ht]
  \centering
  \includegraphics[width=\linewidth]{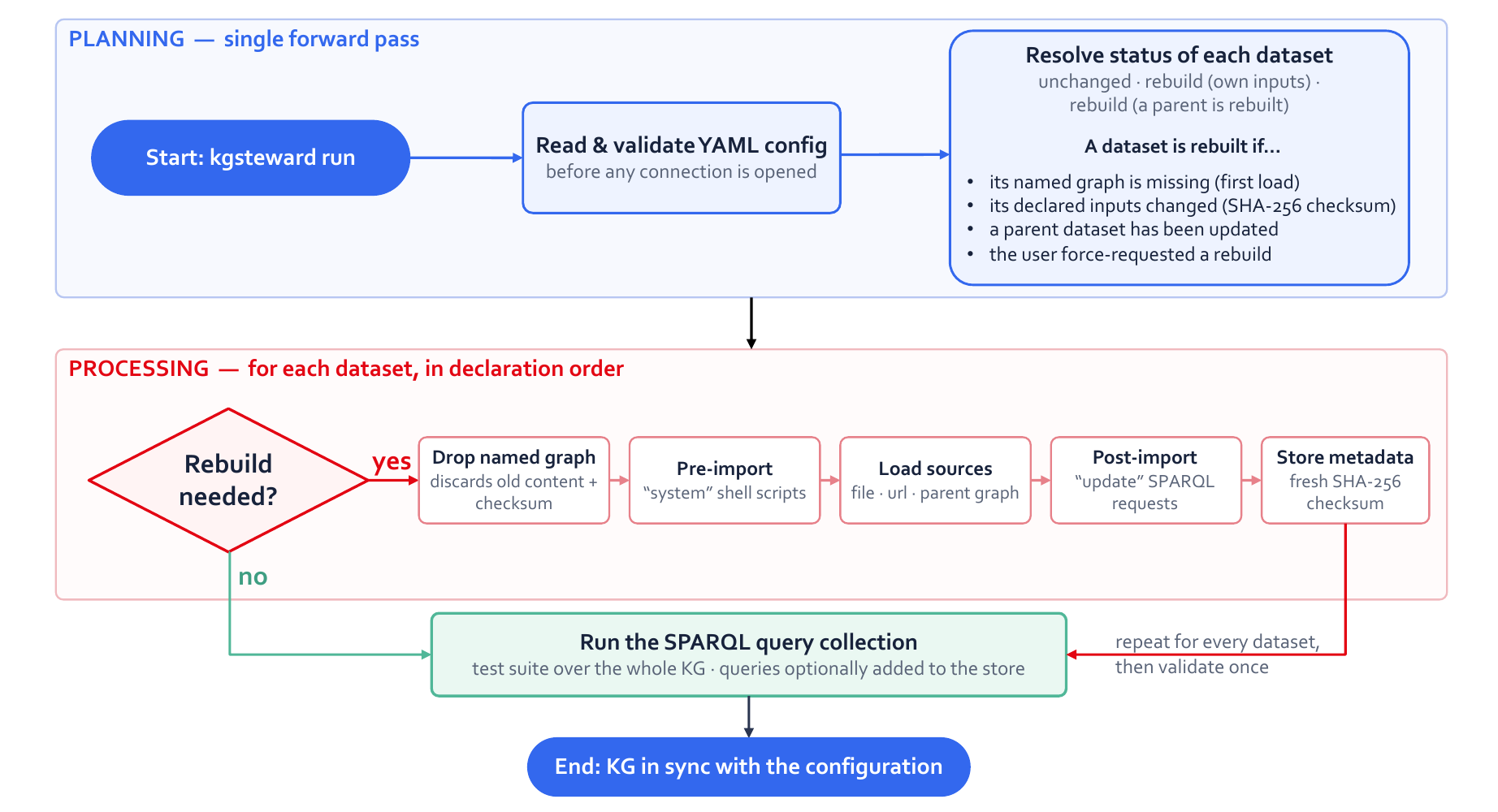}
  \caption{Main steps of a \emph{kgsteward} run.}
  \label{fig:run-steps}
\end{figure}

At each run, \emph{kgsteward} synchronises the local KG with the config file,
reprocessing only the datasets whose inputs have changed
(Figure~\ref{fig:run-steps}). A dataset is rebuilt (re-imported) when:

\begin{itemize}
\item The named graph for the dataset is missing (first load of a dataset).
\item The inputs declared for the dataset have changed since it was last loaded
(detected by checksum).
\item A parent dataset has been updated.
\item The user has force-requested the rebuild of the dataset.
\end{itemize}

\subsection{Configuration file}
\label{sec:config}

\emph{kgsteward} configuration files are written in YAML, and are validated
before any connection to the triplestore is opened, so that a malformed or
incomplete specification is rejected outright rather than part-way through a
load (Figure~\ref{fig:config-example}).

\begin{figure}[!ht]
\begin{lstlisting}
version: kgsteward_yaml_3

server:
  brand            : graphdb
  location         : ${TRIPLESTORE_URL}
  server_config    : graphdb.config.ttl
  repository       : demo_kg
  username         : ${TRIPLESTORE_USER}
  password         : ${TRIPLESTORE_PASSWORD}
  context_base_IRI : http://example.org/context/
  file_loader      : { method: riot_chunk_store }
  url_loader       : { method: sparql_load }

dataset:
  - name : reference_ontology
    url :
      - https://example.org/ontology/v2.ttl

  - name : experimental_data
    file :
      - data/measurements_*.ttl

  - name : integrated_view
    parent :
      - reference_ontology
      - experimental_data
    replace :
      TARGET_GRAPH : ${TARGET_GRAPH_CONTEXT}
    update :
      - sparql/update/link_measurements.rq

queries: !include example_queries.yaml
\end{lstlisting}
  \caption{Example of a minimal \emph{kgsteward} config YAML file.}
  \label{fig:config-example}
\end{figure}

The most important sections of the config files are the following:

\begin{itemize}
\item \textbf{Server}: the local triplestore configuration settings, and in
  particular the triplestore brand.
\item \textbf{Datasets}: the list of datasets from which to build the local KG,
  and the most substantial part of the config file. A dataset's source can be a
  URL, a local file, or a named graph already in the KG to which SPARQL UPDATE
  transformations are then applied.
\item \textbf{Queries}: a set of SPARQL queries used as both test-suite of the
  built KG and documentation.
\end{itemize}

The config file also supports the \texttt{\$\{VARIABLE\}} syntax, which is
expanded to match existing shell variables. This allows for instance to keep
store access credentials out of the config file, a critical aspect when storing
a file in a shared Git repository. Wildcard (glob) expansion is also supported,
allowing a single pattern to match multiple files.

Relative paths are resolved against the directory holding the configuration file
rather than the working directory, so a configuration remains valid wherever it
is invoked from. The content of another YAML file can also be nested and/or
reused via the \texttt{!include} keyword, which proves highly useful for
instance to reuse a SPARQL UPDATE on different datasets.

\subsection{Dataset pre- and post-import instructions}
\label{sec:pre-post}

As an optional feature, a dataset configuration can contain pre- and post-import
instructions. Pre-import instructions (\texttt{system}) are user-supplied shell
commands that are run before a dataset is added to the local KG. A typical
use-case is the decompression of the input files to be loaded.

Post-import instructions (\texttt{update}) are SPARQL UPDATE requests executed
once a dataset's sources have been loaded into the local KG. Their primary
purpose is patching the imported data, which routinely contains errors or lacks
the links needed to connect it to the rest of the KG. Thanks to this post-import
instructions mechanism, user-provided patches are held in the configuration,
under version control, and are automatically re-applied each time a dataset is
updated.

Two further optional mechanisms allow additional customisation of a dataset's
post-import processing: the replacement of variables in update requests
(\texttt{replace}), and a set of hard-coded preset actions (\texttt{special}).

\subsection{Change-detection mechanism}
\label{sec:change-detection}

The decision to update (rebuild) a dataset rests on a SHA-256 checksum computed
on a dataset's declared inputs (Table~\ref{tab:checksum}): its sources, but also
its pre- and post-import instructions. As a result, modifying any of the pre- or
post-import instructions triggers a rebuild of the dataset just as modifying a
source file would. A drawback is that inconsequential edits (e.g.\ adding a
comment line to a pre-import script) change the checksum and trigger a rebuild
of the dataset and all of its children.

Optionally, \texttt{stamp} values can also be set in a dataset's configuration:
these can be e.g.\ local files, remote files, or output of shell commands, and
will contribute to the dataset's checksum (but not used for anything else).

\begin{table}[!ht]
  \caption{Elements contributing to the SHA-256 checksum of a dataset.}
  \label{tab:checksum}
  \begin{tabular}{@{}l>{\raggedright\arraybackslash}p{0.74\linewidth}@{}}
    \toprule
    Config file field & Contribution to checksum \\
    \midrule
    \texttt{context} & Name of the dataset (an IRI). \\
    \texttt{parent}  & Names of all parent datasets. \\
    \texttt{system}  & Content of any pre-import shell script. \\
    \texttt{file}    & Content of every local input file, after wildcard
                       expansion. \\
    \texttt{url}     & URL of an input source. The URL itself, as well as the
                       \enquote{last-modified} and \enquote{content-length}
                       headers returned by an HTTP \enquote{HEAD} request made
                       to the URL contribute to the checksum. \\
    \texttt{stamp}   & Content of the file, or standard output of the shell
                       command. \\
    \texttt{replace} & The substitution variables and their values. \\
    \texttt{update}  & Content of the SPARQL update files. \\
    \texttt{special} & Name of post-import \texttt{special} recipe. \\
    \bottomrule
  \end{tabular}
\end{table}

Since remote data can be heavy to download, the hashing of a dataset's source is
treated differently for local and remote sources. Local source files are read in
full and hashed, so that any change to their content is detected. For remote
sources, the URL is hashed together with the \enquote{last-modified} and
\enquote{content-length} headers of a \enquote{HEAD} request made to the URL.
Detecting a change in a remote source thus relies on the server reporting those
headers faithfully.

The checksum is computed twice in the course of a run: first during planning,
where it is compared to the value stored in the named graph to decide whether
the dataset must be rebuilt; then again after the dataset has been processed,
because pre-import steps (\texttt{system}) can modify the very input files that
are hashed. This second value is the one stored in the named graph for the next
run. Storing the checksum inside the local KG ensures that deleting a named
graph discards the checksum together with the content, so metadata can never
outlive the data it describes.

\subsection{Dependency propagation}
\label{sec:dependencies}

A dataset may declare one or more parents, indicating that its content depends
on the content of the parent datasets. In its config file, \emph{kgsteward}
requires a child dataset to be declared after all of its parents, which allows
the state of the entire configuration to be resolved in a single forward pass.
Each dataset is examined once, in declaration order, and receives a status:
\emph{unchanged} or \emph{to be rebuilt} because its own inputs have changed or
one of its parents is being rebuilt. The datasets are then processed in that
same order, which guarantees that a parent has been updated in full before any
dataset depending on it is recomputed.

An alternative would be to checksum the parent's named graph as stored in the
local KG, and update a child only if that checksum changed, thereby avoiding
rebuilds after inconsequential changes to the parent configuration. Serialising
the content of a named graph is however not
trivial~\cite{hoganCanonicalFormsIsomorphic2017}, and would be overly time
consuming for large graphs.

Finally, a dataset can also be declared as \texttt{frozen}, which excludes it
from any updating. This can e.g.\ be used for datasets that are computationally
intensive to reload.

\subsection{Dataset derivation}
\label{sec:derivation}

Three categories of sources are possible for a dataset: a local file, a remote
server, or a named graph that is added to the local KG by a parent dataset. This
last case creates a so-called \textbf{derived dataset}. A derived dataset
declares no file and no URL as source, but instead supplies one or more SPARQL
update files that are used to compute the derived triples. Import and derivation
are not exclusive, since a dataset may load documents and then transform them in
place.

It is this derivation mechanism that offers the possibility to turn
\emph{kgsteward}'s configuration file from a simple list of sources into a small
data processing pipeline, each stage of which is itself a named graph that can
be inspected, queried and validated independently of the others.

Some limits of the SPARQL language, like the lack of a general recursivity, can
be overcome by creating temporary entities and triples by executing multiple,
sequential, SPARQL updates. Doing so can greatly improve data processing speed
as compared to exporting the data to process it outside the triplestore (i.e.
with a different language than SPARQL).

\subsection{SPARQL query collection and post-import validation tests}
\label{sec:queries}

\emph{kgsteward} configuration files can declare a collection of SPARQL queries.
For practical reasons, these are usually defined in their own files, kept under
version control. Queries can be declared in named groups, organised by theme and
sharing properties in common (Figure~\ref{fig:queries-example}). A group can
also carry a \texttt{test} attribute, giving the expected bounds on the number
of rows a query returns.

\begin{figure}[!ht]
\begin{lstlisting}
queries:
  - name : user_examples   # documented examples, no expectation declared
    file :
      - sparql/examples/*.rq

  - name : coherence       # diagnostic queries: any row returned is a defect
    test :
      max_row_count : 0
    file :
      - sparql/validate/*.rq
\end{lstlisting}
  \caption{Example of a \emph{kgsteward} SPARQL queries configuration file.}
  \label{fig:queries-example}
\end{figure}

The SPARQL query collection serves two important purposes. Its first use is the
validation of the locally generated KG, a test suite for the newly populated
triplestore. The second is documentation: the queries can be uploaded into the
KG to serve purposes such as being integrated in a SPARQL
editor~\cite{emonetUserFriendlySPARQLQuery2026} as examples, or provided to an
MCP (Model Context Protocol) server as AI agent documentation.

SHACL would be an alternative for these post-import validation tasks, but its
implementations are vendor-dependent, and it would require every user to master
an additional language. A collection of SPARQL queries, our current solution, is
more universally useful, as it also serves as a training set for humans and AI
models on how the KG can be queried.

\subsection{Triplestore abstraction}
\label{sec:abstraction}

\emph{kgsteward} was initially developed with GraphDB as its local triplestore
backend, which, despite its limitations for non-commercial usage, is currently
one of the most mature and best documented triplestores. Support for QLever,
rdf4j and Fuseki has since been added, although these backends have not been as
extensively tested on large projects.

The store vendor config specifications are confined to a single block in the
YAML configuration file. Everything else, the datasets, their sources, their
dependencies, the SPARQL transformations and the query collection, is expressed
in terms of named graphs and SPARQL alone, rendering it independent of the store
engine. To remain triplestore vendor-agnostic, SPARQL queries and updates must
stick as much as possible to W3C specifications, with only very limited use of
vendor extensions. For example, here are a few issues we recently faced while
comparing GraphDB and QLever: (1) \texttt{GROUP\_CONCAT} can hardly be made
deterministic in QLever; (2) Collation order is vendor specific and not easily
configurable; (3) the definitions of \texttt{DEFAULT} graph are different.

\emph{kgsteward} also provides instruments to check whether different store
engines are building the same KG. It does so by writing the content of any
dataset, and the result of any query, as a tabular file whose rows are sorted
independently of the order in which the engine returned them. Files obtained
from two engines can then be compared by ordinary textual difference.
Additionally, every SPARQL update issued in the course of a run can be logged in
TSV files with its execution time, so that the performance of different store
engines can be compared.

\subsection{Implementation and availability}
\label{sec:availability}

\emph{kgsteward} is implemented as a CLI program written in Python
($\geq$~3.10), that keeps no state and runs no service of its own: all state
resides in the triplestore it manages. \emph{kgsteward} is open source,
distributed under the GNU General Public License version 3. It is published on
PyPI under the name \enquote{kgsteward}, and can be installed with any standard
Python package manager. The source code is available at
\url{https://github.com/sib-swiss/kgsteward}. The version described in this
document is \texttt{3.5.6}.

From its creation through late 2025 \emph{kgsteward} was developed without AI
contributions. Subsequently, agentic AI was leveraged to drive the project
forward, in particular for exposing curated SPARQL examples to AI agents through
an MCP server and for lowering the entry cost of querying the knowledge graph
for non-specialist users.

\section{First use case: managing the Sinergia Wolfender knowledge graph as a
  shared resource}
\label{sec:usecase-sinergia}

\subsection{Project context}
\label{sec:sinergia-context}

The knowledge graph (KG) of this first use case was developed within the SNSF
Sinergia project \enquote{An in silico and chemo-biological approach to identify
anti-infective and pro-metabolic natural products} (grant no. CRSII5\_189921/1),
a consortium bringing together the SIB Swiss Institute of Bioinformatics, the
University of Geneva, ETH Zurich and the Pierre Fabre laboratory. The project
investigated a uniquely biodiverse library of more than 17,000 plant-derived
natural extracts, combining mass-spectrometry metabolomics, cheminformatics
predictions, plant taxonomy, and bioassay results to prioritise bioactive
natural products against targets such as tuberculosis and
obesity~\cite{gaudrySampleCentricKnowledgeDrivenComputational2024,%
allardOpenReusableAnnotated2023}.

The KG evolved in stages: an initial pilot graph established the RDF and
semantic-integration blueprint; this prototype was then refactored into ENPKG,
the Experimental Natural Products Knowledge Graph, centred on an open-source
dataset of 1,600 chemodiverse plant extracts comprising more than 1,000,000
LC-HRMS/MS spectra, 2,665 spectrally matched compounds and full taxonomic
annotation~\cite{allardOpenReusableAnnotated2023}. ENPKG was first applied in a
sample-centric, knowledge-driven screen against trypanosomatid parasites,
enabling the rapid dereplication of known and previously unknown
anti-\emph{Trypanosoma cruzi}
compounds~\cite{gaudrySampleCentricKnowledgeDrivenComputational2024}.

Beyond the initial construction of the KG, the primary role of \emph{kgsteward}
was to support the long-term management of a heterogeneous and continuously
evolving resource within a distributed consortium. Rather than producing a
single static graph, the project used the KG as a shared data-management
substrate integrating public reference data, private consortium datasets,
in-house predictions, and external resources.

Contributors, primarily PhD students and postdoctoral researchers with diverse
backgrounds and levels of technical expertise, were free to use their preferred
programming languages and tools to generate RDF data, provided they adhered to
vocabularies and semantic conventions developed collaboratively within the
consortium. Both the data and their semantics evolved progressively throughout
the project.

Although configurations and part of the data are shared through an
access-controlled Git repository, governance remains distributed. Each partner
can deploy, extend, and re-synchronise its own KG instance independently,
incorporating new datasets without relying on a central administrator. The KG
itself, together with its controlled evolution, is therefore the object
maintained by \emph{kgsteward}. Consequently, the framework should be evaluated
primarily by its ability to support the distributed stewardship and sustainable
evolution of a shared knowledge resource, rather than by any single scientific
outcome.

\subsection{Vocabularies}
\label{sec:sinergia-vocabularies}

Contributors were initially given considerable freedom to use existing
vocabularies or propose new ones, provided that namespace clashes were avoided.
As expertise increased, these vocabularies could be reviewed and refined. This
approach differs from the common recommendation to maximise reuse of existing
vocabularies from the outset. While reuse promotes interoperability, it can also
introduce semantic ambiguities when RDF from multiple sources is aggregated. For
example, \texttt{skos:prefLabel}, which is intended to be unique, may become
problematic after merging independently curated datasets.

Our recommendation was therefore to document ad hoc vocabularies through
explicit mappings to established semantic constructs such as
\texttt{rdfs:subClassOf}, \texttt{rdfs:subPropertyOf},
\texttt{owl:equivalentProperty}, and \texttt{owl:sameAs}. We believe that these
relationships provide sufficient semantic context for both human users and
agentic AI systems. The flexibility offered by \emph{kgsteward}, particularly
its use of SPARQL Update operations to transform imported RDF, greatly
facilitated such semantic experimentation and gradual harmonisation.

\subsection{Contents of the knowledge graph}
\label{sec:sinergia-contents}

The graph is organised around the concepts the project uses to structure its
data. The collection of Pierre Fabre samples is partitioned into sub-collections
named \enquote{PFcode}. Each PFcode is a collection batch annotated with plant
taxonomy and plant part, and resolves to physical samples that may be
fractionated or extracted into derived samples. Analytical work is captured as
\enquote{assay batches}, each a coherent set of results produced by one operator
at a given time with a defined instrument, software and protocol. Assay batches
span three modalities: LC-MS chemistry, bioassays (anti-infective screening,
antitrypanosomal testing, pro-metabolic assays, and Wnt-signalling assays) and
in-silico prioritisation. Individual measurements, purified chemical entities
(each with its molecular identity and per-assay bioactivity interpretation), and
the persons and provenance behind every dataset complete the model. Reflecting
this layering, the ENPKG vocabulary describes the chemistry (LC-MS) layer, while
a project-wide vocabulary completed with vocabularies DCAT and PROV-O covers
collections, samples, assays, readouts and provenance.

The final knowledge graph comprises 644,544,052 triples (5.5 GB of compressed
Trig), deposited on Zenodo~\cite{pagniKnowledgeGraphSilico2025}; an earlier
conference snapshot had reported a work-in-progress figure of about 200 million
triples~\cite{burdetApplicationRDFFramework2024}. The project's data summary
(sinergiawolfender.org/data\_summary.html) is generated with an R script
querying the graph using SPARQL~\cite{englerSparqlr}: it inventories every
integrated dataset, collections, assay batches, measurements, chemical entities,
contributors, in searchable tables that link into the GraphDB resource browser.
Each assay batch carries its operator, date, protocol reference, original files
and an explicit licence, and people and compounds are identified by ORCID and
InChIKey. The summary thus supplies the metadata FAIR reuse requires, while
serving as documentation that cannot drift from the data, and as an entry point
for users who do not write SPARQL. The validation suite maintained for this
graph comprises tens of SPARQL queries, which are executed as post-import tests
after every build and, at the same time, serve as usage examples for end users
and AI agents (Section~\ref{sec:queries}).

Working from this integrated resource, the consortium assembled a focused
library of rotenoid analogues active against \emph{T.~cruzi}, 41 isolated
compounds including 11 rotenoids, seven of them previously unreported, with
deguelin and rotenone as the potent
hits~\cite{gaudryEfficientConstitutionLibrary2025}. Coupling the graph with a
stringent 3R infection model (\emph{Dictyostelium discoideum} infected with
\emph{Mycobacterium marinum}, a surrogate for macrophage /
\emph{M.~tuberculosis} infection) prioritised stilbene-rich extracts and led to
the isolation of 11 stilbene oligomers from \emph{Gnetum edule}, six of them new
stereoisomers, the most active being ($-$)-gnetuhainin M
(IC$_{50}$~$\approx$~22~$\mu$M)~\cite{kirchhofferPrioritizationNovelAntiinfective2025}.
More broadly, the graph records 42 purified compounds profiled directly in the
anti-infective assay, drawn from six plant species. Beyond the project’s
original anti-infective and pro-metabolic scope, natural extracts from the same
Pierre Fabre collection were also screened for antiviral activity: an approach
combining high-throughput screening with metabolomics led to the discovery of
anti-respiratory syncytial virus (RSV) compounds from plant
extracts~\cite{haemmerliInnovativeApproachAntirespiratory2026}.

Complementary chemodiverse profiling of the collection --- for example 76
extracts spanning 36 species of the Celastraceae family, rich in triterpenoids
and yielding known bioactives such as celastrol, maytansine and triptolide ---
further extended the annotated reference space available for graph-based
discovery~\cite{Quiros2024Comprehensive}.

\subsection{From graph to agentic access: MetaboT}
\label{sec:sinergia-metabot}

ENPKG subsequently served as the foundation for MetaboT, an LLM-based
multi-agent system that translates natural-language questions into SPARQL
queries over the graph. MetaboT couples six coordinated agents with dedicated
chemical, taxon and target resolvers; on a 50-question benchmark it answered
83.7\% of questions correctly, compared with 8.2\% for a standalone GPT-4o
model, the improvement being most pronounced for high-complexity queries (79\%
versus 0\%)~\cite{bekbergenovaMetaboTLLMbasedMultiAgent2025}. This illustrates
concretely how the availability of curated SPARQL examples, the same ones
\emph{kgsteward} uses for validation and exposes via its MCP server, lowers the
entry cost of a knowledge graph for non-specialist users.

\section{Second use case: the ReconXKG metabolic knowledge graph}
\label{sec:usecase-reconxkg}

The second use case is an unpublished work in progress, and we therefore present
it as an ongoing illustration rather than a completed, extensively demonstrated
result. Whereas the Sinergia graph integrated \emph{experimental data} produced
within one consortium, ReconXKG integrates and reconciles many \emph{public
reference resources} for human metabolism and one currently private metabolic
reconstruction VMH2. ReconXKG is developed within the EU Horizon Recon4IMD
project (Reconstruction and Computational Modelling for Inherited Metabolic
Diseases) as an open workspace supporting the extension of the human metabolic
network reconstruction Recon3 towards
Recon4~\cite{brunkRecon3DEnablesThreedimensional2018}. It is currently deployed
as an internal consortium resource before becoming a publicly accessible
endpoint.

\subsection{A different data-management problem}
\label{sec:reconxkg-problem}

\begin{figure}[!ht]
  \centering
  \includegraphics[width=\linewidth]{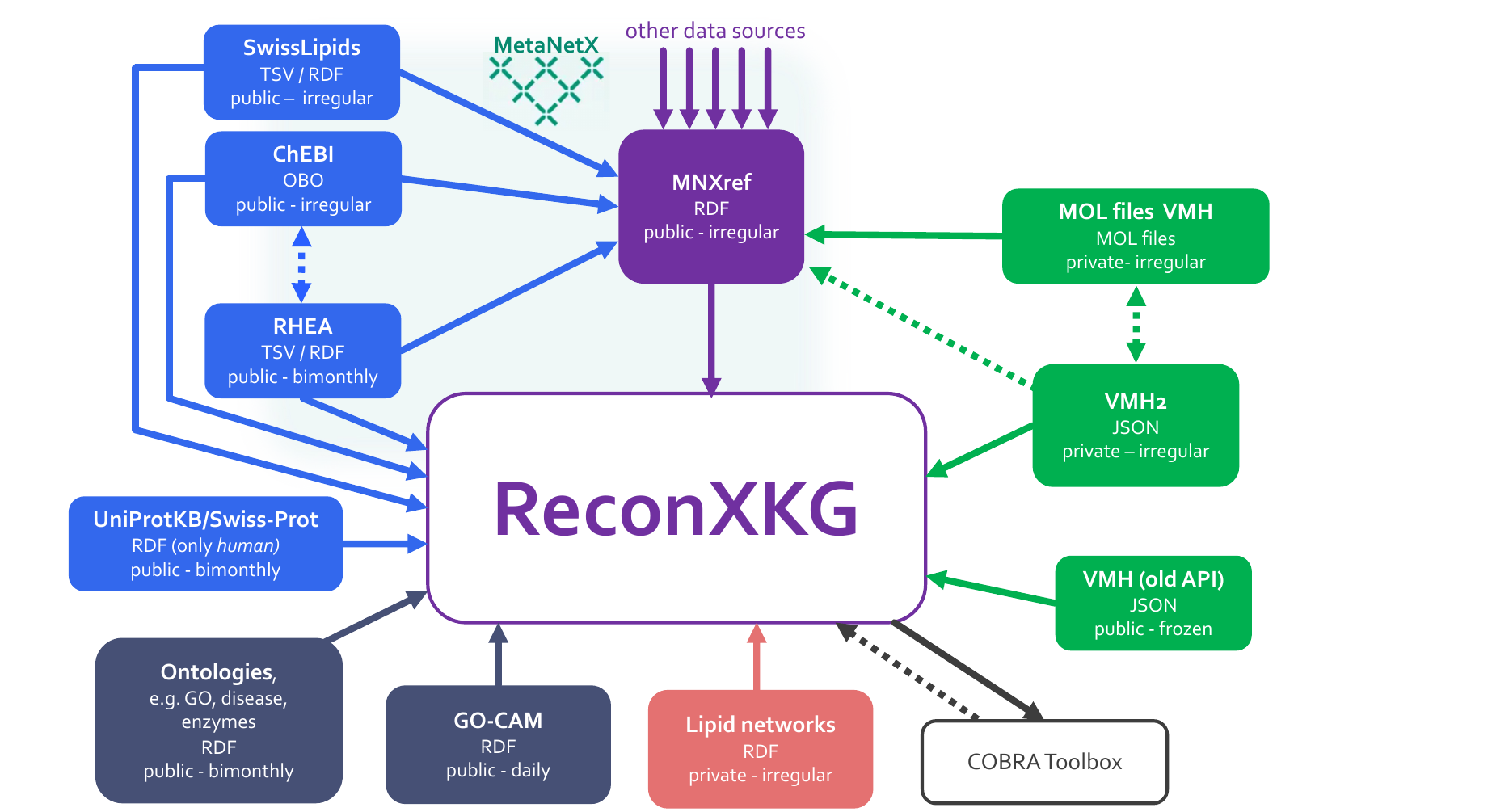}
  \caption{ReconXKG reconciles different resources pertaining to human
    metabolism. \emph{kgsteward} assembles and harmonises the RDF linking
    layer via SPARQL updates.}
  \label{fig:reconxkg}
\end{figure}

ReconXKG aggregates metabolites, reactions and gene/protein/reaction
relationships from complementary resources: the Virtual Metabolic Human (VMH /
VMH2)~\cite{noronhaVirtualMetabolicHuman2019a}, the MetaNetX reconciliation
namespace~\cite{morettiMetaNetXBridgeMetabolic2026}, and the
UniProt~\cite{theuniprotconsortiumUniProtUniversalProtein2025} and
Rhea~\cite{bansalRheaReactionKnowledgebase2022} knowledge bases
(Figure~\ref{fig:reconxkg}). It also annotates these resources with links to
further databases such as ChEBI, SwissLipids and various ontologies. In contrast
with the Sinergia Wolfender project, the difficulty here is not capturing
freshly produced data but aligning mature resources that differ in terminology,
data structure and curation practice. The reconciliation occurs at two different
levels, at the metabolite level and at the metabolic network level. MetaNetX
proposes a clustering of molecular entities as metabolites that make sense in
the context of biochemistry, thus solving the problem of individual IRI
reconciliation. Those metabolites are substrates or products of biochemical
reactions that constitute metabolic networks. Metabolic networks are complex
data structures available from multiple sources. However, each source represents
these networks using completely different RDF structures and vocabularies,
making direct comparison and integration within a KG very difficult.

The idea was to convert each of these sources into a common data model by
leveraging on SPARQL UPDATE transformations. We pushed the idea even further and
produced multiple versions of a given metabolic network, as the different
partners in the project have different favoured namespaces for metabolites,
reactions and enzymes. This eventually enables the systematic comparison,
integration, and harmonisation of multiple metabolic networks in multiple
versions within a single KG. The execution of these SPARQL updates is fully
automated by \emph{kgsteward}, as are the updates of the different sources.
Altogether, ReconXKG is becoming a resource about human metabolism with a degree
of integration and a level of complexity rarely matched. Currently, the primary
applications of this KG are quality control and reconciliation of the different
resources within the consortium. It will be progressively transformed into a
repository of knowledge about human cellular metabolism. Because it is powered
by a single YAML configuration, its replication elsewhere should serve as a
basis for future applications, including the analysis of embargoed data.

\section{Related work: positioning \emph{kgsteward} among existing tools}
\label{sec:related}

\emph{kgsteward} owes much to previous experience we gained by using the DVC
Data Version Control~\cite{ruslankuprieievDVCDataVersion2026}, a data versioning
and pipeline management tool built on top of Git that enables reproducible
analyses and remote storage of large datasets.

\emph{kgsteward}'s conceptually most direct competitor is Data2Services (d2s), a
framework that transforms structured data into RDF and loads it into a
triplestore through containerised
workflows~\cite{instituteofdatascienceData2ServicesD2sFramework}. In the same
\enquote{RDF ETL pipeline} vein, LinkedPipes ETL provides configurable
components to extract, transform and load
RDF~\cite{klimekLinkedPipesETLEvolved2016}. In the ontology world, the OBO
Foundry's Ontology Development Kit (ODK) plays an analogous role, reproducible
builds driven by configuration and Makefiles, but is aimed at the publication of
ontologies rather than at the content of a live
endpoint~\cite{matentzogluOntologyDevelopmentKit2022}. Relative to these,
\emph{kgsteward} is deliberately simpler: it is driven by a single configuration
file, oriented towards named graphs, and focused on the reproducible rebuild of
an existing SPARQL endpoint, whereas d2s and LinkedPipes are heavier and more
general frameworks.

Much of what \emph{kgsteward} automates can also be done manually using the
native tools shipped by the triplestores themselves: the Apache Jena
command-line utilities (\texttt{tdbloader}, \texttt{s-put}, and the
SPARQL-over-HTTP SOH scripts) together with RDF Delta for versioning and
patching Fuseki
datasets~\cite{theapachesoftwarefoundationApacheJenaCommandline}; the preload /
ImportRDF utilities and the Workbench of
GraphDB~\cite{ontotextOntotextGraphDBRDF}; \texttt{stardog-admin} on the Stardog
side; and the Virtuoso bulk loader (\texttt{isql}). These operate at a lower
level and are specific to one vendor. In practice it is most often against such
home-made scripts that \emph{kgsteward} positions itself, replacing them with a
single declarative, store-agnostic and version-controlled configuration.

Finally, for biomedical knowledge graphs built as property graphs rather than
strict RDF, BioCypher offers a unifying construction
framework~\cite{lobentanzerDemocratizingKnowledgeRepresentation2023}, as do the
KG-Hub / KGX / Koza tools of the Monarch Initiative
ecosystem~\cite{kornKozaKozaHubBorninteroperable2025}. These alternative tools
are relevant primarily when the choice between RDF and property graphs remains
open.

\section{Discussion}
\label{sec:discussion}

\subsection{Simplicity as a requirement, not a convenience}
\label{sec:discussion-simplicity}

Collaborative research projects are not staffed by RDF specialists. In the
Sinergia Wolfender project the contributors were mainly PhD students and
postdoctoral researchers in chemistry and biology, and their willingness to
produce data at all set the ceiling on what the graph could contain. For the
contributor, \emph{kgsteward} asks for nothing beyond a file placed in a Git
repository: RDF may be generated in whatever language suits, and the vocabulary
conventions are progressively agreed within the consortium rather than a
constraint enforced at start. Any partner can therefore host and rebuild the
graph without operating infrastructure or acquiring specialist skill. This
enables a distributed governance and allows staff to experiment with RDF, with
limited risk of clash as long as vocabulary namespaces are controlled.

\subsection{Evolving projects and embargoed data}
\label{sec:discussion-evolving}

Neither project began with a stable schema. Instead, the data model was revised
repeatedly as understanding improved. A configuration-driven rebuild is what
makes such revisions affordable: because the entire graph is derived from a
versioned configuration, changing the model means editing a file and re-running
\emph{kgsteward}, while checksum-driven selective re-import keeps the cost
proportional to what actually changed. Semantic experimentation, in particular,
becomes something a project can afford to do more than once.

\subsection{SPARQL UPDATE, an underused integration primitive}
\label{sec:discussion-sparql-update}

The lesson we did not anticipate concerns SPARQL UPDATE. A large and growing
share of the reference data these projects depend on is already published as
RDF. For such sources the integration problem is not \enquote{how do I turn this
into triples}, which is what the mapping-language literature addresses, but
\enquote{how do I reconcile triples that are already there}. SPARQL UPDATE
answers that question directly and inside the store: no parser to write, no
intermediate serialisation, no round trip out of the database.

ReconXKG pushes this furthest. Each source describes metabolic networks using
its own RDF structures and vocabularies; SPARQL UPDATE queries convert them into
a common data model and, beyond that, into several parallel versions of the same
network, one per partner's preferred vocabulary, making the networks
systematically comparable. Known limitations of the language, notably the
absence of general recursion, can be worked around by chaining updates through
temporary triples. We therefore believe SPARQL UPDATE deserves more attention as
a first-class data-transformation language. Held under version control,
replayed at every build and producing named graphs that can be inspected and
validated independently, it turns a configuration into an auditable pipeline
expressed entirely in a language the project already needs to know.

\begin{acknowledgments}
  We thank the Swiss National Science Foundation for financial support,
  specifically through the SNSF Sinergia project \enquote{An in silico and
  chemo-biological approach to identify anti-infective and pro-metabolic natural
  products} (CRSII5\_189921/1) and the Bridge-Discovery grant
  \enquote{Identification of antimicrobial and anti-virulence lead compounds
  from biodiverse microorganisms through a platform integrating metabolomics and
  innovative bioassays} (211759). This research was also supported by Recon4IMD,
  co-funded by the European Union's Horizon Europe Framework Programme
  (101080997), the Swiss State Secretariat for Education, Research and
  Innovation (23.00232), and by United Kingdom Research and Innovation (10083717
  \& 10080153). We also thank the Jena/Fuseki and QLever teams for their support
  and for their excellent open-source tools, and all members of the Sinergia
  Wolfender consortium who contributed data to the knowledge graph.
\end{acknowledgments}

\section*{Declaration on Generative AI}

During the preparation of this manuscript, the author(s) used Claude Opus 5 in
order to: assist in drafting content, assist in improving writing style. After
using this tool, the author(s) reviewed and edited the content as needed and
take(s) full responsibility for the publication's content.

\bibliography{kgsteward-swat4hcls}

\end{document}